\documentclass[aps,pra,reprint,twocolumn,nofootinbib,a4paper,floatfix,10pt]{revtex4-1}

\usepackage{color}
\usepackage{bm}
\usepackage{amsmath,amssymb}
\usepackage{amsfonts}
\usepackage{dsfont}
\usepackage{graphicx}
\usepackage{float}
\usepackage{subfigure}
\usepackage{verbatim}
\usepackage{amsfonts}
\usepackage{dcolumn}
\usepackage{bm}
\usepackage{color}
\usepackage[dvipsnames]{xcolor}
\usepackage{listings}
\usepackage[colorlinks=true,urlcolor=Blue,citecolor=Blue]{hyperref}
\usepackage{mathrsfs}
\usepackage{filecontents}

\usepackage{ntheorem}

\newcommand{\bra}[1]{\mbox{$\langle #1 |$}}
\newcommand{\ket}[1]{\mbox{$| #1 \rangle$}}
\newcommand{\tr}{\mbox{tr}}

\newcommand{\mycirc}[1][black]{\Large\textcolor{#1}{\ensuremath\bullet}}

\begin{document}

\title{Quantifying entanglement in two-mode Gaussian states}
\author{Spyros Tserkis}  \email{s.tserkis@uq.edu.au}
\affiliation{Centre for Quantum Computation and Communication Technology, School of Mathematics and Physics, University of Queensland, St Lucia, Queensland 4072, Australia}
\author{Timothy C. Ralph}  \email{ralph@physics.uq.edu.au}
\affiliation{Centre for Quantum Computation and Communication Technology, School of Mathematics and Physics, University of Queensland, St Lucia, Queensland 4072, Australia}
\date{\today}

\begin{abstract}
Entangled two-mode Gaussian states are a key resource for quantum information technologies such as teleportation, quantum cryptography and quantum computation, so quantification of Gaussian entanglement is an important problem. Entanglement of formation is unanimously considered a proper measure of quantum correlations, but for arbitrary two-mode Gaussian states no analytical form is currently known. In contrast, logarithmic negativity is a measure straightforward to calculate and so has been adopted by most researchers, even though it is a less faithful quantifier. In this work, we derive an analytical lower bound for entanglement of formation of generic two-mode Gaussian states, which becomes tight for symmetric states and for states with balanced correlations. We define simple expressions for entanglement of formation in physically relevant situations and use these to illustrate the problematic behavior of logarithmic negativity, which can lead to spurious conclusions.
\end{abstract}

\maketitle

Entanglement is a non-classical physical property, emerging from the quantum mechanical superposition principle. Theoretically, it can be described as the inability to separate a global quantum state of a composite system into a product of individual subsystems. Experimentally, it is manifested as the correlations of the observables of different subsystems, which cannot be classically reproduced.

In order to quantify entanglement of bipartite systems, we employ the axiomatic theory of entanglement measures \cite{Plenio.Virmani.B.14,Horodecki.et.al.RMP.09}, where an entanglement measure, $\mathcal{E} $, should satisfy the following postulates: i) $\mathcal{E} $ vanishes on separable states, and ii) $\mathcal{E} $ does not increase on average under local operations and classical communication (strong monotonicity). Besides the above postulates, there are several other mathematical properties that it is desirable for $\mathcal{E}$ to satisfy, such as additivity, strong superadditivity, convexity and asymptotic continuity. 

For pure states, entropy of entanglement is the \textit{bona fide} measure of quantum correlations, defined as $\mathcal{E}(\ket{\psi}){:=}\mathcal{S} (\tr_B \ket{\psi}\bra{\psi})$, where $\mathcal{S}(\rho){:=}{-}\tr(\rho\log_2 \rho)$ is the von Neumann entropy, and $\tr_B$ denotes the partial trace over subsystem $B$ \cite{Bennett.et.al.PRL.96}. For mixed states, entanglement can be measured via different quantifiers, which, in general, do not coincide with each other. One of them is entanglement of formation, defined as the convex-roof extension of the von Neumann entropy, $\mathcal{E}_F(\rho){:=}{\inf} {\{\sum_i}{p_i} {\mathcal{S}}{({\tr_B}{\ket{\psi_i}}{\bra{\psi_i}})}\}$, where the infimum is taken over all ensembles $\{{p_i}{,}{\psi_i}\}$ of $\rho{:=}{\sum_i}{p_i}{\ket{\psi_i}}{\bra{\psi_i}}$ \cite{Bennett.DiVincenzo.et.al.PRA.96}. Specifically, for two-mode Gaussian states, where $\mathcal{E}_F$ has been proven to be additive \cite{Marian.Marian.PRL.08} (and thus strongly superadditive as well \cite{Pomeransky.PRA.03}), it coincides with the entanglement cost, $\displaystyle {\mathcal{E}_C(\rho)}{:=}{\lim_{n \rightarrow \infty}} {\mathcal{E}_F}{(\rho^{\otimes n})}/n$ \cite{Hayden.Horodecki.Terhal.JPAMG.01}. For a given state $\rho$, entanglement cost has a clear operational meaning, since it quantifies the minimum entanglement needed (cost of quantum resources) to produce $\rho$ \cite{Hayden.Horodecki.Terhal.JPAMG.01}, which is of great importance in quantum technologies. In discrete-variables (DV) bipartite systems, an explicit form of entanglement of formation has been found for generic states (qubits) \cite{Wootters.PRL.98}, while in the continuous-variables (CV) regime, and specifically for two-mode Gaussian systems, there are only two families where the entanglement of formation can be analytically calculated: a) for symmetric states \cite{Giedke.et.al.PRL.03}, and b) for non-symmetric extremal (maximally and minimally) entangled states for fixed global and local purities (GMEMS/ GLEMS) \cite{Adesso.Serafini.Illuminati.PRA.04,Adesso.Illuminati.PRA.05,Akbari-Kourbolagh.Alijanzadeh-Boura.QIP.15,Giovannetti.et.al.NP.14}. An explicit form of the measure for arbitrary two-mode Gaussian states is yet considered an open problem. 

The inability to define entanglement of formation through an explicit closed form for arbitrary states, led researches to use other, more easily computable measures. Specifically, in two-mode Gaussian systems the most widely used quantifier is the logarithmic negativity, ${\mathcal{E}_N}{(\rho)}{:=}{\log_2}\| \tilde{\rho} \|$, where $\tilde{\rho}$ denotes the partially transposed density matrix $\rho$, and ${\| x \|}{:=}\tr\sqrt{x^{\dag}x}$ is the trace norm \cite{Zyczkowski.et.al.PRA.98,Vidal.Werner.PRA.02,Plenio.PRL.05}. However, unlike $\mathcal{E}_F$, $\mathcal{E}_N$ does not satisfy convexity, asymptotic continuity and strong superadditivity \cite{Plenio.Virmani.B.14,Horodecki.et.al.RMP.09,Wolf.Giedke.Cirac.PRL.06}. Asymptotic continuity and strong superadditivity are requirements for an entanglement measure to satisfy the widely accepted extremality of Gaussian states, i.e., for a given covariance matrix the entanglement is minimized by Gaussian states \cite{Wolf.Giedke.Cirac.PRL.06,Adesso.PRA.09}. Logarithmic negativity not only fails to satisfy those requirements, but counterexamples have also been found, showing that $\mathcal{E}_N$ can actually defy the extremality of Gaussian states, leading to an overestimation of entanglement \cite{Wolf.Giedke.Cirac.PRL.06}. Furthermore, since logarithmic negativity is not asymptotically continuous, it does not reduce to the entropy of entanglement on all pure states \cite{Plenio.Virmani.B.14}, which is the reason why it is usually referred to as a monotone, instead of a measure.

In this work we provide a clear physical interpretation of the entanglement of formation and we derive an analytical lower bound of it for arbitrary two-mode Gaussian states, which saturates for symmetric states and for states with balanced correlations. For the rest of the states, the bound provides a measure of necessary correlations needed to construct the state, closely approximating the exact value (computed numerically) of the entanglement of formation. Our approach leads to simple exact expressions for the $\mathcal{E}_F$ of two-mode squeezed states after passage through typical communication channels, that we use to illustrate significant qualitative differences between $\mathcal{E}_N$ and $\mathcal{E}_F$.

We begin by briefly reviewing two-mode Gaussian states \cite{Weedbrook.et.al.RVP.12,Adesso.Illuminati.JPAMT.07}. Any two-mode state can be fully described by a covariance matrix (assuming for simplicity that its mean value is zero), that in standard form \cite{Duan.et.al.PRL.00,Simon.PRL.00}, is written as
\begin{equation}
\boldsymbol{\sigma}^{\text{sf}}=\begin{bmatrix}
\boldsymbol{A} & \boldsymbol{C} \\
\boldsymbol{C} & \boldsymbol{B} 
\end{bmatrix}\,,
\label{cov}
\end{equation}
which is a real and positive definite matrix, with $\boldsymbol{A}{=}\text{diag}(a{,}a)$, $\boldsymbol{B}{=}\text{diag}(b{,}b)$, and $\boldsymbol{C}{=}\text{diag}(c_1{,}c_2)$. Its elements are proportional to the second-order moments of the quadrature field operators, $\hat{x}_j{:=}\hat{a}_j{+}\hat{a}^{\dag}_j$ and $\hat{p}_j{:=}i(\hat{a}^{\dag}_j{-}\hat{a}_j)$, where $\hat{a}_j$ and $\hat{a}^{\dag}_j$ are the annihilation and creation operators, respectively, with $[\hat{a}_i{,}{\hat{a}^{\dag}_j}]{=}\delta_{ij}$. In CV optical systems entanglement is manifested by the correlations of the field operators $\hat{x}$ and $\hat{p}$, and it is typically created by pumping a nonlinear crystal in a non-degenerate optical parametric amplifier. This process is described by a Gaussian unitary known as the two-mode squeezing operator defined as ${S_2(r)}{:=}{\exp[r(\hat{a}\hat{b}{-}\hat{a}^{\dag}\hat{b}^{\dag})/2]}$, where $r \in \mathbb{R}$ is the squeezing parameter. By applying $S_2(r)$ to a couple of vacua, we obtain a pure state called the two-mode squeezed vacuum (TMSV), with $a{=}b{=}\frac{1+\chi^2}{1-\chi^2}$ and $c_1{=}{-}c_2{=}\frac{2\chi}{1-\chi^2}$, where $\chi{=}\tanh r \in [0{,}1)$.

For any covariance matrix $\boldsymbol{\sigma}$, there exists a symplectic transformation $S$, such that $\boldsymbol{\sigma}{=}S \mathbf{\boldsymbol{\nu}}S^{T}$, with $\boldsymbol{\nu}{=}\nu_-\mathds{1} {\oplus} \nu_+\mathds{1}$, where $1 {\leq} \nu_- {\leq} \nu_+$. The quantities $\nu_i$ are called symplectic eigenvalues \cite{Vidal.Werner.PRA.02}. The necessary and sufficient separability criterion for a two-mode Gaussian state $\boldsymbol{\sigma}$ has been shown to be the positivity of the partial transposed state $\boldsymbol{\tilde{\sigma}}$ \cite{Duan.et.al.PRL.00,Simon.PRL.00,Giovannetti.et.al.PRA.03}. This is equivalent to checking the condition ${\tilde{\nu}_-} {\geq} 1$ \cite{Adesso.Serafini.Illuminati.PRA.04}, where ${\tilde{\nu}_-}$ is the lowest symplectic eigenvalue of $\boldsymbol{\tilde{\sigma}}$.

\begin{figure}[b]
\centering
  \includegraphics[width=\columnwidth]{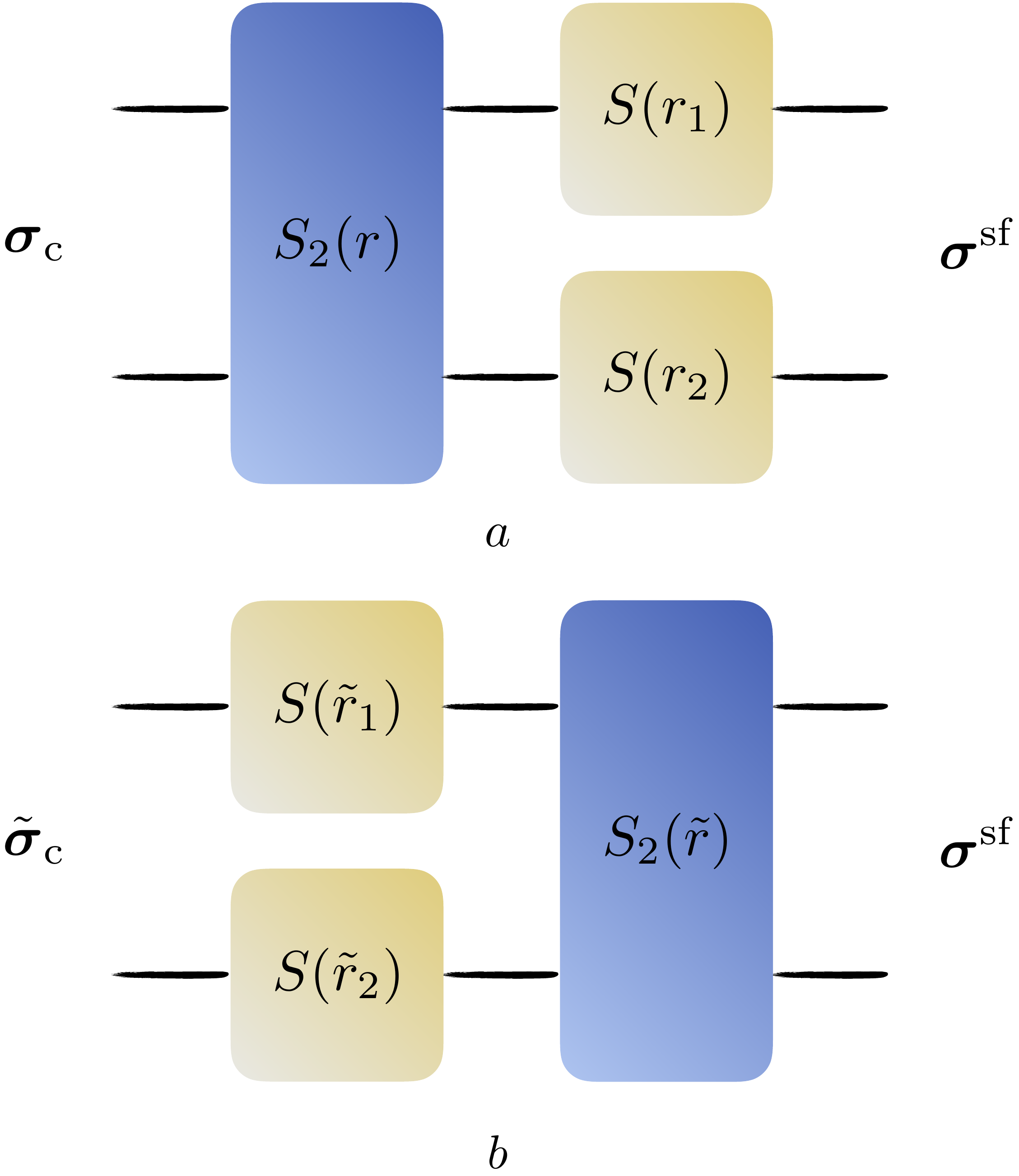}
    \caption{ \small State decompositions. Any state, $\boldsymbol{\sigma}^{\text{sf}}$, can be constructed by applying a sequence of a) two-mode squeezing $S_2(r)$ followed by local squeezing $S(r_i)$ on a classical state $\boldsymbol{\sigma}_{\text{c}}$ or reversely, b) local squeezing $S(\tilde{r}_i)$ followed by two-mode squeezing $S_2(\tilde{r})$ on a classical state $\tilde{\boldsymbol{\sigma}}_{\text{c}}$.}
    \label{fig1}
\end{figure}

Any state $\boldsymbol{\sigma}^{\text{sf}}$ can be decomposed (proof can be found in the Appendix \ref{ap1}) as
\begin{equation}
\boldsymbol{\sigma}^{\text{sf}}=L(r_1,r_2)S_2(r) \boldsymbol{\sigma}_{\text{c}} S_2^T(r) L^T(r_1,r_2)\,,
\label{dec1}
\end{equation}
where $L(r_1,r_2){=}S(r_1){\oplus}S(r_2)$ is the local squeezing operation ${S(r)}{:=}{\exp[r(\hat{a}^2{-}\hat{a}^{\dag 2})/2]}$ on each mode, and $ \boldsymbol{\sigma}_{\text{c}}{\geq}\mathds{1}$ is a classical state (see Fig.~\ref{fig1}a). We call \textit{optimum} the decomposition with the least amount of two-mode squeezing, $r_o$, i.e., $\boldsymbol{\sigma}^{\text{sf}}{=} L(r_{1_o},r_{2_o})S_2(r_o) \boldsymbol{\sigma}_{\text{c}_o}S_2^T(r_o)L^T(r_{1_o},r_{2_o})$. Gaussian entanglement of formation \cite{Wolf.et.al.PRA.04}, which has been proven to be equal to the general entanglement of formation in two-mode Gaussian systems \cite{Akbari-Kourbolagh.Alijanzadeh-Boura.QIP.15}, is equal to the von Neumann entropy \cite{Holevo.Sohma.Hirota.PRA.99} of the pure state with the minimum amount of two-mode squeezing, $\boldsymbol{\sigma}_p(r_o){=}L(r_{1_o},r_{2_o})S_2(r_o) \mathds{1} S_2^T(r_o) L^T(r_{1_o},r_{2_o})$ (with the corresponding symplectic eigenvalue, ${\tilde{\nu}_{o_-}}{=}e^{-2 r_o}$), and thus ${\mathcal{E}_F} {(\boldsymbol{\sigma})}{:=}{\mathcal{S}}{[\boldsymbol{\sigma}_p(r_o)]}$, so we have
\begin{equation}
{\mathcal{E}_F} {(\boldsymbol{\sigma})}={\cosh^2} {r_o} {\log_2} {( {\cosh^2} {r_o}){-}{\sinh^2} {r_o} {\log_2} ( {\sinh^2} {r_o})}\,.
\label{EoF}
\end{equation}
Thus, \textit{entanglement of formation quantifies the minimum amount of two-mode squeezing needed to prepare an entangled state starting from a classical one}. The optimum decomposition, and consequently $r_o$, cannot in general be found analytically \cite{Wolf.et.al.PRA.04,Marian.Marian.PRL.08,Akbari-Kourbolagh.Alijanzadeh-Boura.QIP.15,Ivan.Simon.arXiv.08} 

Another way to decompose a state is the following
\begin{equation}
\boldsymbol{\sigma}^{\text{sf}}=S_2(\tilde{r})L(\tilde{r}_1,\tilde{r}_2) \tilde{\boldsymbol{\sigma}}_{\text{c}} L^T(\tilde{r}_1,\tilde{r}_2)S_2^T(\tilde{r})\,,
\label{dec2}
\end{equation}
since we can always disentangle a state by anti-squeezing it and then apply the corresponding local squeezing to make the separable state classical, i.e., $ \tilde{\boldsymbol{\sigma}}_{\text{c}}{\geq}\mathds{1}$ (see Fig.~\ref{fig1}b). In order to make a state separable we have to solve the inequality $\tilde{\nu}_-\left[ S_2(-\tilde{r}) \boldsymbol{\sigma}^{\text{sf}} S_2^T(-\tilde{r}) \right] {\geq} 1$, which is satisfied for a range of $\tilde{r}_- {\leq} \tilde{r} {\leq} \tilde{r}_+$, with
\begin{equation}
\tilde{r}_{\pm}=\frac{1}{2} \ln \sqrt{\frac{\kappa\pm\sqrt{\kappa^2- \lambda_+ \lambda_-}}{\lambda_-}}\,,
\label{lowerbound}
\end{equation}
where we have set  $\kappa{= }2(\det \boldsymbol{\sigma} {+} 1){-}(a{-}b)^2$ and $\lambda_{\pm}{=}\det \boldsymbol{A}{+} \det \boldsymbol{B}{-}2\det \boldsymbol{C}{+}2[(a b {-} c_1 c_2){\pm} (c_1{-}c_2)(a{+}b)]$. The physical meaning of $\tilde{r}_-$ is that it quantifies the minimum amount of two-mode squeezing needed to disentangle a state (in its standard form). For symmetric states, i.e., $a{=}b$, and for states with balanced correlations, i.e., $c_1{=}{-}c_2$, we have $\tilde{r}_-{=}r_o$, but in general $\tilde{r}_-{\leq}r_o$ (the proof of that statement can be found in the Appendix \ref{ap2}), and thus we have a lower bound of the entanglement of formation
\begin{equation}
{\tilde{\mathcal{E}}_F} {(\boldsymbol{\sigma})}:={\mathcal{S}}{[\boldsymbol{\sigma}_p(\tilde{r}_-)]}\leq {\mathcal{S}}{[\boldsymbol{\sigma}_p(r_o)]}:={\mathcal{E}_F} {(\boldsymbol{\sigma})}\,.
\end{equation}
We note that for fixed global and local purities, the more imbalanced the correlations the larger the deviation of the lower bound ${\tilde{\mathcal{E}}_F} {(\boldsymbol{\sigma})}$ from the real value ${\mathcal{E}_F} {(\boldsymbol{\sigma})}$.

\begin{figure}[b]
\centering
  \includegraphics[width=\columnwidth]{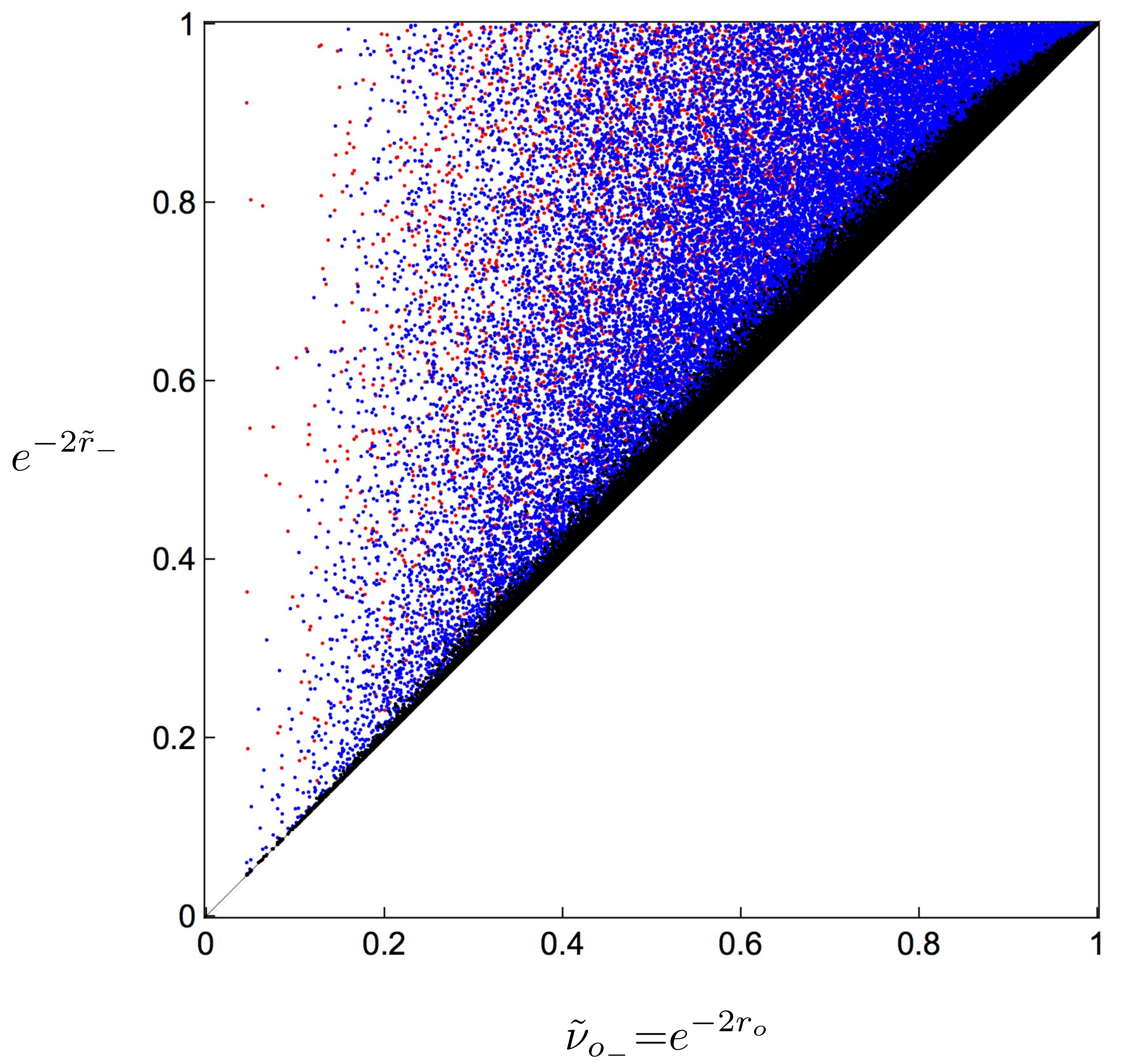}
    \caption{ \small Lower bound for entanglement of formation. We plot with black dots {\mycirc} the optimum symplectic eigenvalue ${\tilde{\nu}_{o_-}}{=}e^{-2 r_o}$ against the corresponding value based on $\tilde{r}_-$, i.e., $e^{-2 \tilde{r}_-}$, for randomly generated states. The symplectic eigenvalue is a bounded value $\in (0,1]$, which shows that a) ${\tilde{\mathcal{E}}_F} {(\boldsymbol{\sigma})}{\leq}{\mathcal{E}_F} {(\boldsymbol{\sigma})}$ and b) that the bound is also tight for separable and infinite entangled states. We also depict with blue {\mycirc[blue]} \cite{Rigolin.Escobar.PRA.04} and red {\mycirc[red]} \cite{Nicacio.Oliveira.PRA.14} dots the corresponding values we get from the previously known lower bounds. The closer the dots are to the diagonal the smaller the deviation from the real value of entanglement. It is clear that our bound is, on average, tighter that previous bounds.}
  \label{fig2}
\end{figure}

In Fig.~\ref{fig2} we compare the entanglement of formation (calculated numerically) and its lower bound for randomly generated states. The significant progress over the previously known lower bounds of the measure derived in \cite{Rigolin.Escobar.PRA.04} and \cite{Nicacio.Oliveira.PRA.14} is also depicted. As we see, former lower bounds deviate significantly from the real value, and sometimes even imply separability for an entangled state. 

For many quantum communication protocols, Gaussian channels describe the decoherence introduced by the environment on a quantum state, and represent the basic models of communication lines such as optical fibres \cite{Weedbrook.et.al.RVP.12}. Let us assume that a single mode of a TMSV state, i.e., $a{=}b{=}\frac{1+\chi^2}{1-\chi^2}$ and $c_1{=}{-}c_2{=}\frac{2\chi}{1-\chi^2}$, with $\chi{=}\tanh r \in [0,1)$, is sent through a Gaussian channel. One-mode Gaussian channels can be defined as the transformation of the covariance matrix of the mode $\boldsymbol{\gamma}$, i.e., $\boldsymbol{\gamma} \rightarrow \mathcal{U} \boldsymbol{\gamma} \, \mathcal{U}^T {+} \mathcal{V}$ \cite{Weedbrook.et.al.RVP.12}.  Typically, these channels are phase invariant and so produce states with balanced correlations that saturate the lower bound, i.e,  $\tilde{r}_-{=}r_o$. The value of $r_o$, derived from $\tilde{r}_-$ in Eq.~\ref{lowerbound}, for three fundamental Gaussian channels is presented below:
\begin{itemize}
\item Lossy channel, $\mathcal{L}(\tau)$, is defined as $\mathcal{U} {=} \sqrt{\tau} \mathds{1}$ and $\mathcal{V}{=}(1{-}\tau) \mathds{1}$, with transmissivity $0{\leq} \tau {\leq} 1$. Thus we have
\begin{equation*}
r_o=\frac{1}{2} \ln \frac{ 1 +\chi \sqrt{\tau}}{1 - \chi \sqrt{\tau}}\,.
\end{equation*}
\item Amplifier channel, $\mathcal{A}(\tau)$, is defined as $\mathcal{U} {=} \sqrt{\tau} \mathds{1}$ and $\mathcal{V}{=}(\tau{-}1) \mathds{1}$, with transmissivity $\tau {\geq} 1$. Eq.~\ref{lowerbound} takes the form
\begin{equation*}
r_o=\frac{1}{2} \ln \frac{\sqrt{\tau}+ \chi }{\sqrt{\tau} -\chi }\,.
\end{equation*}
\item Classical noise channel, $\mathcal{C}(v)$, is defined as $\mathcal{U} {=}\mathds{1}$ and $\mathcal{V}{=}v \mathds{1}$, with $v {\geq} 0$. The optimum squeezing parameter for $0 {\leq} v {\leq} 2$ is
\begin{equation*}
r_o=\frac{1}{2} \ln \frac{2+v+\chi(2-v)}{2+v+\chi(v-2)}\,,
\end{equation*}
while for $v{>}2$, $r_o$ vanishes, i.e., entanglement-breaking bound.
\end{itemize}
The deterministic upper bound of entanglement for a channel, i.e., the amount of entanglement assuming an infinitely squeezed state is sent through the same channel \cite{Ulanov.et.al.NP.15}, is reached for $\chi \rightarrow 1$. This bound allows us to investigate physical limits, like the calculation of the maximum possible amount of quantum correlations that can possibly exist after a specific decohering channel.

As mentioned before, besides entanglement of formation, other quantifiers have also been used to compute entanglement for these kinds of states, so it would be interesting to give a direct comparison with the most popular of those (due to its computability), i.e., the logarithmic negativity, which is defined, for two-mode Gaussian states, as ${\mathcal{E}_N} {(\boldsymbol{\sigma})} {:=} {\max} {\left[0{,} {-}{\log_2} {{\tilde{\nu}_-}} \right]}$ \cite{Zyczkowski.et.al.PRA.98,Vidal.Werner.PRA.02,Plenio.PRL.05,Adesso.Serafini.Illuminati.PRA.04}. In order to have a clear operational meaning of this monotone, we can define the generalized \textit{EPR} correlations $\hat{u}{=} \frac{\hat{x}_1-g_x \hat{x}_2}{\sqrt{1+{g_x}^2}}$ and  $\hat{v}{=} \frac{\hat{p}_1+g_p \hat{p}_2}{\sqrt{1+{g_p}^2}}$, where $g_x,g_p \in \mathbb{R}$ are experimentally variable gains. For those operators the separability criterion \cite{Giovannetti.et.al.PRA.03} takes the form $\beta{=}\frac{V_x V_p}{(1+g_x {g_p})^2} {\geq} 1$, with $V_x{=}\langle ( \hat{x}_1 {-}g_x \hat{x}_2) ^2 \rangle$ and $V_p{=}\langle ( \hat{p}_1 {+} {g_p} \hat{p}_2) ^2 \rangle$ being the conditional variances. For $\beta{<}1$ we have an entangled state, and its minimum value $\beta_-$ is equal to ${{\tilde{\nu}_-}}^2$ \cite{He.Gong.Reid.PRL.15}. We should note that this equality, i.e., $\beta_-{=}{{\tilde{\nu}_-}}^2$, holds for any two-mode Gaussian state. So, \textit{logarithmic negativity quantifies the maximum possible violation of the separability criterion}. 

\begin{figure}[t]
\centering
  \includegraphics[width=\columnwidth]{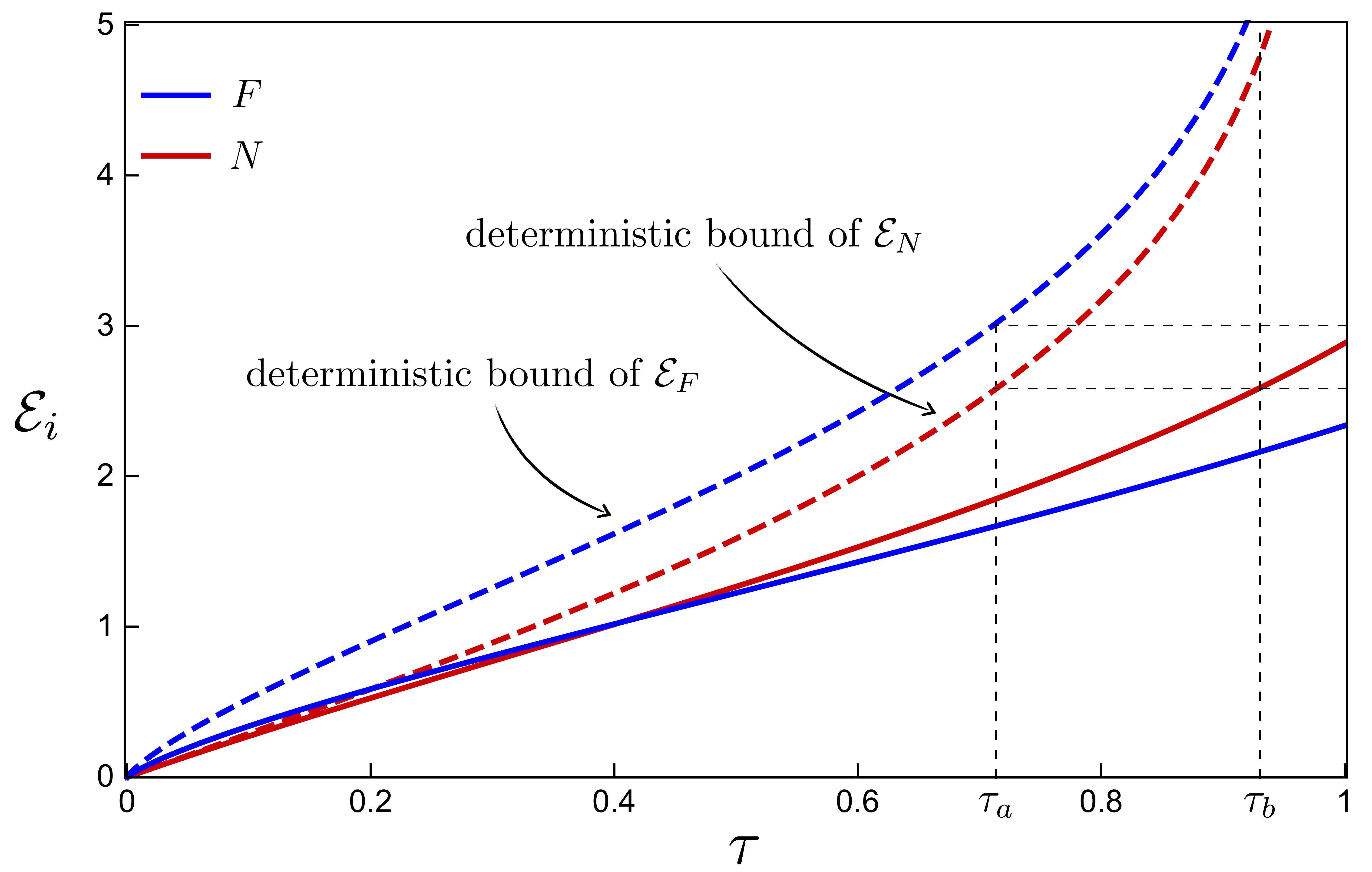}
  \caption{\small Comparison between entanglement of formation and logarithmic negativity. Assuming a TMSV state with $r{=}1$ is sent through a lossy channel of transmissivity $0 {\leq}\tau {\leq} 1$, we compare the two measures. The deterministic bounds, i.e., the amount of entanglement assuming an infinitely squeezed state is sent through the same channel, are also depicted with the corresponding dashed lines, since they provide further insight regarding the qualitative differences between entanglement of formation and logarithmic negativity. The deterministic bound for logarithmic negativity can be found in ref.~\cite{Ulanov.et.al.NP.15}. Specifically, for logarithmic negativity, the deterministic bound of a state with transmissivity value $\tau_a$, can also be reached by sending the squeezed state $(r{=}1)$ through a channel of transmissivity $\tau_b$, with $\tau_b {>} \tau_a$. However, in contrast, entanglement of formation predicts that we cannot reach the deterministic bound with a squeezed state $(r{=}1)$ regardless of how much we raise the transmissivity. This is a critical difference, since the two quantifiers disagree on whether a physical upper bound has been reached or not.}
  \label{fig3}
\end{figure}

Logarithmic negativity is, in general, not directly related to the squeezing of the state, which is a major drawback, since squeezing is considered as the resource of the quantum correlations in the system and is, experimentally, the primary figure of merit. Furthermore, in Fig.~\ref{fig3} it is apparent that $\mathcal{E}_N$ fails, in general, to satisfy the extremality of entanglement cost (which coincides with entanglement of formation in these systems), i.e., $\mathcal{E}_i {\leq} \mathcal{E}_C$ \cite{Donald.Horodecki.Rudolph.JMP.02}, which was expected since logarithmic negativity is not asymptotically continuous. That results in an inconsistent behavior of $\mathcal{E}_N$, which, for finite squeezing, can either be an upper or lower bound of $\mathcal{E}_F$, depending on the channel that the state is sent through. A specific example of how $\mathcal{E}_N$ can lead to a qualitatively different evaluation of the entanglement sent through a physically relevant channel compared to $\mathcal{E}_F$ is shown in Fig.~\ref{fig3}. To sum up, logarithmic negativity is a quantifier widely used in the literature, since it has the merit of being analytically computable in various quantum systems, but from a information-theoretic point of view is inferior to entanglement of formation.

In conclusion, we have found a lower bound of entanglement of formation which serves as the minimum amount of correlations needed to construct a state. This lower bound is tight for symmetric states and for states with balanced correlations, while it deviates from the real value for states with asymmetric correlations. The deviation, though, is relatively small, which practically makes this lower bound an analytical approximation of the entanglement of formation for experimental purposes. We also showed via physical examples that this measure should be favored compared to logarithmic negativity. We also introduced an alternative interpretation of the measure in Gaussian systems, proving that entanglement of formation is intrinsically related to the amount of anti-squeezing needed to disentangle a state up to the point that the state becomes classical, which might also be helpful for the quantification of entanglement of several other families of states, e.g., multipartite Gaussian or non-Gaussian states. 

We thank Saleh Rahimi-Keshari for motivating discussions and Gerardo Adesso for the enlightening comments on our first version of the manuscript. The research is supported by the Australian Research Council (ARC) under the Centre of Excellence for Quantum Computation and Communication Technology (CE110001027).

\clearpage

\onecolumngrid
\appendix

\section{State decomposition}
\label{ap1}

Any two-mode Gaussian state can be written in the standard form as \cite{Marian.Marian.PRL.08,Ivan.Simon.arXiv.08,Akbari-Kourbolagh.Alijanzadeh-Boura.QIP.15}

\begin{equation}
\boldsymbol{\sigma}^{\text{sf}}=L(r_1,r_2)\left[\boldsymbol{\sigma}_p^{\text{sf}}(r)+ \boldsymbol{\varphi}\right]L^T(r_1,r_2) \,,
\end{equation}
where $L(r_1,r_2){=}S(r_1){\oplus}S(r_2)$ is the local squeezing operation ${S(r)}{:=}{\exp[r(\hat{a}^2{-}\hat{a}^{\dag 2})/2]}$ on each mode, and $\boldsymbol{\varphi}$ is a positive semidefinite matrix. So, we have

\begin{equation}
\boldsymbol{\sigma}^{\text{sf}}=L(r_1,r_2)\left[\underbrace{S_2(r)\mathds{1}S_2^T(r)}_{\boldsymbol{\sigma}_p^{\text{sf}}(r)}+S_2(r)\underbrace{S_2(-r)\boldsymbol{\varphi}S_2^T(-r)}_{\boldsymbol{\theta}}S_2^T(r)\right]L^T(r_1,r_2)\,.
\end{equation}

Since $\boldsymbol{\varphi}$ has a structure identical to a covariance matrix, but not necessarily in the standard form, i.e., 

\begin{equation}
\boldsymbol{\varphi}=\begin{bmatrix}
n_1& 0 & d_1 & 0 \\
0 & n_2 & 0 & d_2 \\
d_1& 0 & m_1 & 0 \\
0 & d_2 & 0 & m_2 
\end{bmatrix}\,,
\end{equation}
then $\boldsymbol{\theta}{=}S_2(-r)\boldsymbol{\varphi}S_2^T(-r)$ is also in the same form as $\boldsymbol{\varphi}$, and thus a Hermitian matrix, so, based on Wigner's theorem \cite{Wigner.CJM.63}, we know that $\boldsymbol{\theta} {\geq} 0$. So, we can write

\begin{equation}
\boldsymbol{\sigma}^{\text{sf}}=L(r_1,r_2)\left[S_2(r)\{\underbrace{\mathds{1}+\boldsymbol{\theta}}_{\boldsymbol{\sigma}_{\text{c}}}\}S_2^T(r)\right]L^T(r_1,r_2)\,,
\end{equation}
but $\mathds{1}{+}\boldsymbol{\theta}$ can always represent a classical state, $\boldsymbol{\sigma}_{\text{c}}$, where $\boldsymbol{\theta}$ is interpreted as the random correlated displacements applied on a couple of vacua, and thus we have 

\begin{equation}
\boldsymbol{\sigma}^{\text{sf}}=L(r_1,r_2)S_2(r) \boldsymbol{\sigma}_{\text{c}} S_2^T(r) L^T(r_1,r_2)\,.
\label{apdec1}
\end{equation}

\section{Lower Bound}
\label{ap2}

Any state can be decomposed as
\begin{equation}
\boldsymbol{\sigma}^{\text{sf}}=S_2(\tilde{r})L(\tilde{r}_1,\tilde{r}_2) \tilde{\boldsymbol{\sigma}}_{\text{c}} L^T(\tilde{r}_1,\tilde{r}_2)S_2^T(\tilde{r})\,,
\label{apdec2}
\end{equation}
since we can always disentangle a state by anti-squeezing it and then apply the corresponding local squeezing to make the separable state classical, i.e., $ \tilde{\boldsymbol{\sigma}}_{\text{c}}{\geq}\mathds{1}$. In order to make a state separable we have to solve the inequality $\tilde{\nu}_-\left[ S_2(-\tilde{r}) \boldsymbol{\sigma}^{\text{sf}} S_2^T(-\tilde{r}) \right] {\geq} 1$, which is satisfied for a range of $\tilde{r}_- {\leq} \tilde{r} {\leq} \tilde{r}_+$, with
\begin{equation}
\tilde{r}_{\pm}=\frac{1}{2} \ln \sqrt{\frac{\kappa\pm\sqrt{\kappa^2- \lambda_+ \lambda_-}}{\lambda_-}}\,,
\end{equation}
where we have set  $\kappa{= }2(\det \boldsymbol{\sigma} {+} 1){-}(a{-}b)^2$ and $\lambda_{\pm}{=}\det \boldsymbol{A}{+} \det \boldsymbol{B}{-}2\det \boldsymbol{C}{+}2[(a b {-} c_1 c_2){\pm} (c_1{-}c_2)(a{+}b)]$. Given a $\tilde{r}$ which disentangles $\boldsymbol{\sigma}^{\text{sf}}$, the local squeezing parameters $\tilde{r}_1$ and $\tilde{r}_2$ needed to remove any non-classicality are

\begin{align}
\tilde{r}_1=&\frac{1}{2} \ln \Bigg[ \frac{1}{2 \left(2 \sinh (2 r_o) \left(a b c_2-c_2 c_1^2+c_1\right)+(a+b) \cosh (2 r_o) \left(a b-c_1^2-1\right)-(a-b) \left(a b-c_1^2+1\right)\right)} \, \times \nonumber \\
& \times \Bigg( -2 +2 \left(a b-c_1^2\right) \left(a b-c_2^2\right)-2 (a-b) ((a+b) \cosh (2 r_o)+(c_2-c_1) \sinh (2 r_o)) \, - \nonumber \\
&-\sqrt{2 \left(a^2 \left(b^2-1\right)-a b \left(c_1^2+c_2^2\right)-b^2+c_1 c_2 (c_1 c_2-2)+1\right)} \, \times \nonumber \\
& \times \sqrt{a^2 \left(2 b^2-1\right)-2 a b \left(c_1^2+c_2^2-1\right)+2 (a+b) (c_1-c_2) \sinh (4 r_o)-\cosh (4 r_o) \left((a+b)^2-4 c_1 c_2\right)-b^2+2 c_1^2 c_2^2+2} \Bigg) \Bigg] \,,
\end{align}
and

\begin{align}
\tilde{r}_2=&\frac{1}{2} \ln \Bigg[ \frac{1}{2 \left(2 \sinh (2 r_o) \left(a b c_2-c_2c_1^2+c_1\right)+(a+b) \cosh (2 r_o) \left(a b-c_1^2-1\right)+(a-b) \left(a b-c_1^2+1\right)\right)} \, \times \nonumber \\
& \times \Bigg( -2+2 \left(a b-c_1^2\right) \left(a b-c_2^2\right)+2 (a-b) ((a+b) \cosh (2 r_o)+(c_2-c_1) \sinh (2 r_o)) \, + \nonumber \\
&+\sqrt{2 \left(a^2 \left(b^2-1\right)-a b \left(c_1^2+c_2^2\right)-b^2+c_1 c_2 (c_1 c_2-2)+1\right)} \, \times \nonumber \\
& \times \sqrt{a^2 \left(2 b^2-1\right)-2 a b \left(c_1^2+c_2^2-1\right)+2 (a+b) (c_1-c_2) \sinh (4 r_o)-\cosh (4 r_o) \left((a+b)^2-4 c_1 c_2\right)-b^2+2 c_1^2 c_2^2+2} \Bigg) \Bigg] \,.
\end{align}

The entanglement needed to construct a state for an arbitrary decomposition of this form is equivalent to the entanglement of the corresponding pure state 

\begin{equation}
\boldsymbol{\sigma}_p=S_2(\tilde{r})L(\tilde{r}_1,\tilde{r}_2) \mathds{1} L^T(\tilde{r}_1,\tilde{r}_2)S_2^T(\tilde{r})\,,
\end{equation}
but the covariance matrix of this pure state is always identical to the covariance matrix constructed in the following way

\begin{equation}
\boldsymbol{\sigma}_p=L(r'_1,r'_2)S_2(r') \mathds{1} S_2^T(r') L^T(r'_1,r'_2)\,,
\end{equation}
where 

\begin{equation}
r'(\tilde{r},\tilde{r}_1,\tilde{r}_2)=\cosh ^{-1}\left(\frac{1}{2} \sqrt{e^{-\tilde{r}_1-\tilde{r}_2} \sqrt{\cosh (2 \tilde{r}) \left(e^{2 \tilde{r}_1}+e^{2 \tilde{r}_2}\right)+e^{2 \tilde{r}_1}-e^{2 \tilde{r}_2}} \sqrt{\cosh (2 \tilde{r}) \left(e^{2 \tilde{r}_1}+e^{2 \tilde{r}_2}\right)-e^{2 \tilde{r}_1}+e^{2 \tilde{r}_2}}+2}\right) \,,
\end{equation}

\begin{equation}
r'_1(\tilde{r},\tilde{r}_1,\tilde{r}_2)= \log \left(\frac{e^{\frac{\tilde{r}_1+\tilde{r}_2}{2}} \sqrt[4]{e^{2 \tilde{r}_1} \cosh ^2(\tilde{r})+e^{2 \tilde{r}_2} \sinh ^2(\tilde{r})}}{\sqrt[4]{e^{2 \tilde{r}_1} \sinh ^2(\tilde{r})+e^{2 \tilde{r}_2} \cosh ^2(\tilde{r})}}\right) \,,
\end{equation} 
and
\begin{align}
r'_2(\tilde{r},\tilde{r}_1,\tilde{r}_2)=& \frac{1}{\left(e^{2 \tilde{r}_1}+e^{2 \tilde{r}_2}\right) \sqrt[4]{e^{2 \tilde{r}_1} \cosh ^2(\tilde{r})+e^{2 \tilde{r}_2} \sinh ^2(\tilde{r})}} \, \times \nonumber \\
& \times \log \Bigg( \text{csch}(\tilde{r}) \text{sech}(\tilde{r}) e^{\frac{3 (\tilde{r}_1+\tilde{r}_2)}{2}} \sqrt[4]{e^{2 \tilde{r}_1} \sinh ^2(\tilde{r})+e^{2 \tilde{r}_2} \cosh ^2(\tilde{r})} \, \times \nonumber \\
& \times \sqrt{e^{-\tilde{r}_1-\tilde{r}_2} \sqrt{e^{2 \tilde{r}_1} \sinh ^2(\tilde{r})+e^{2 \tilde{r}_2} \cosh ^2(\tilde{r})} \sqrt{e^{2 \tilde{r}_1} \cosh ^2(\tilde{r})+e^{2 \tilde{r}_2} \sinh ^2(\tilde{r})}-1} \, \times \nonumber \\
& \times \sqrt{e^{-\tilde{r}_1-\tilde{r}_2} \sqrt{e^{2 \tilde{r}_1} \sinh ^2(\tilde{r})+e^{2 \tilde{r}_2} \cosh ^2(\tilde{r})} \sqrt{e^{2 \tilde{r}_1} \cosh ^2(\tilde{r})+e^{2 \tilde{r}_2} \sinh ^2(\tilde{r})}+1} \Bigg) \,.
\end{align}

Let assume that we have the optimum decomposition for the entanglement of formation, i.e., $\boldsymbol{\sigma}^{\text{sf}}{=} L(r_{1_o},r_{2_o})S_2(r_o) \boldsymbol{\sigma}_{{\text{c}}_o} S_2^T(r_o)L^T(r_{1_o},r_{2_o})$, which corresponds to $\boldsymbol{\sigma}^{\text{sf}}{=}S_2(\tilde{r}_o)L(\tilde{r}_{1_o},\tilde{r}_{2_o}) \tilde{\boldsymbol{\sigma}}_{{\text{c}}_o}L^T(\tilde{r}_{1_o},\tilde{r}_{2_o})S_2^T(\tilde{r}_o)$, with $\tilde{r}_-{\leq}\tilde{r}_o{\leq}\tilde{r}_+$. We know that $r_o$ must be a function of $r'$, i.e., $r_o{=}r'(\tilde{r}_o,\tilde{r}_{1_o},\tilde{r}_{2_o})$. It is straightforward to prove that  $r'(\tilde{r},\tilde{r}_1,\tilde{r}_2){\geq} r'(\tilde{r},\tilde{r}_1{=}\tilde{r}_2)$, since $\frac{\partial r'}{\partial \tilde{r}_1}{=}\frac{\partial r'}{\partial \tilde{r}_2}{=}0\Rightarrow r_1{=}r_2$ and $\frac{\partial^2 r'}{\partial \tilde{r}^2_1}{\geq}0$, $\frac{\partial^2 r'}{\partial \tilde{r}^2_2}{\geq}0$ for any $\tilde{r}{>}0$.  So, for the case of $\tilde{r}_{1_o}{=}\tilde{r}_{2_o}{=}0$, $r'(\tilde{r}_o,\tilde{r}_{1_o},\tilde{r}_{2_o}){\geq} r'(\tilde{r}_o,\tilde{r}_{1_o}{=}\tilde{r}_{2_o}{=}0)$ should hold as well. It is apparent that $r'(\tilde{r}_o,\tilde{r}_{1_o}{=}\tilde{r}_{2_o}{=}0){=}\tilde{r}_o$, and thus 

\begin{equation}
\tilde{r}_- \leq r_o \Rightarrow {\tilde{\mathcal{E}}_F} {(\boldsymbol{\sigma})}:={\mathcal{S}}{[\boldsymbol{\sigma}_p(\tilde{r}_-)]}\leq {\mathcal{S}}{[\boldsymbol{\sigma}_p(r_o)]}:={\mathcal{E}_F} {(\boldsymbol{\sigma})} \,,
\end{equation}
where $ {\tilde{\mathcal{E}}_F} {(\boldsymbol{\sigma})}$ is the lower bound of entanglement of formation. The reason why this lower bound is tight for balanced states is because for those states the local squeezing parameters of the optimum decomposition are found to be $r_1{=}r_2{=}0$ \cite{Marian.Marian.PRL.08,Ivan.Simon.arXiv.08,Akbari-Kourbolagh.Alijanzadeh-Boura.QIP.15}, and thus the two decompositions, i.e., Eq.~\ref{apdec1} and Eq.~\ref{apdec2}, coincide. For symmetric states, where the local squeezing parameters of the optimal decomposition are $r_1{=}r_2{=}\sqrt{\frac{a+c_2}{a-c_1}}$ \cite{Wolf.et.al.PRA.04,Marian.Marian.PRL.08,Ivan.Simon.arXiv.08,Akbari-Kourbolagh.Alijanzadeh-Boura.QIP.15}, the bound is tight since the operation $L(r_1,r_2)S_2(r) S_2^T(r) L^T(r_1,r_2)$ is identical to $S_2(r) L(r_1,r_2) L^T(r_1,r_2) S_2^T(r)$ for $r{=}r_1{=}r_2$, so again the two decompositions (Eq.~\ref{apdec1} and Eq.~\ref{apdec2}) coincide.

\end{document}